\documentclass[lettersize,journal]{IEEEtran}
\usepackage{amsmath,amsfonts}
\usepackage{algorithmic}
\usepackage{algorithm}
\usepackage{array}
\usepackage[caption=false,font=normalsize,labelfont=sf,textfont=sf]{subfig}
\usepackage{textcomp}
\usepackage{stfloats}
\usepackage{url}
\usepackage{verbatim}
\usepackage{graphicx}
\usepackage{cite}
\begin{document}

\title{Towards Quantum Resilience: \\ Data-Driven Migration Strategy Design}

\author{Ozan Çetin, Emil Huseynov, Nahid Aliyev~\IEEEmembership{}
\thanks{O. Çetin, E. Huseynov, and N. Aliyev are with the Department of Computer Engineering, Istanbul Technical University, Istanbul, Turkey}
\thanks{Manuscript received April 22, 2025}}

\markboth{Journal of \LaTeX\ Class Files,~Vol.~XX, No.~X, April~2025}%
{Shell \MakeLowercase{\textit{et al.}}: A Sample Article Using IEEEtran.cls for IEEE Journals}


\maketitle

\begin{abstract}
The advancements in quantum computing are a threat to classical cryptographic systems. The traditional cryptographic methods that utilize factorization-based or discrete-logarithm-based algorithms, such as RSA and ECC, are some of these. This paper thoroughly investigates the vulnerabilities of traditional cryptographic methods against quantum attacks and provides a decision-support framework to help organizations in recommending mitigation plans and determining appropriate transition strategies to post-quantum cryptography. A semi-synthetic dataset, consisting of key features such as key size, network complexity, and sensitivity levels, is crafted, with each configuration labeled according to its recommended mitigation plan. Using decision tree and random forest models, a classifier is trained to recommend appropriate mitigation/transition plans such as continuous monitoring, scheduled transitions, and immediate hybrid implementation. The proposed approach introduces a data-driven and dynamic solution for organizations to assess the scale of the migration, specifying a structured roadmap toward quantum resilience. The results highlight important features that influence strategy decisions and support actionable recommendations for cryptographic modernization based on system context.
\end{abstract}


\section{Introduction}
\IEEEPARstart{O}{ver} the past few decades, public-key cryptography has emerged as a critical element of digital environment security. Numerous protocols, such as RSA (Rivest-Shamir-Adleman), Elliptic Curve Cryptography (ECC), and Diffie-Hellman (DH) are widely used for a variety of purposes, including user authentication, sensitive data protection, and secure communications in a variety of systems, ranging from e-commerce and internet banking platforms to governmental and military infrastructures \cite{sood2024}. These approaches rely on computational complexity of mathematical problems that are thought to be impossible to solve with traditional computers, such as discrete logarithm problem and integer factorization.  \\ 

However, the advancements in quantum computing are an emerging threat against these traditional methods, and they present a paradigm shift with subtle implications for cryptography. Particularly, Shor's algorithm, developed in 1994, proved that strong quantum computers can efficiently factor large integers and compute discrete logarithms \cite{zhou2023}. Classical brute-force attacks scale exponentially proportional to the key size. However, Shor's algorithm can solve this problem in polynomial time, drastically reducing the effort to compromise encrypted systems.  \\ 

The large-scale quantum computers that are capable of executing Shor's algorithm on cryptographically relevant key sizes have not yet been realized, but a variety of initiatives by big tech companies like IBM, Google, and other research institutions signals that this risk is no longer in theory \cite{memon2024}. There is steady progress in quantum hardware, and the estimate of a real threat is likely to emerge within the next 10 to 20 years, according to the experts.  \\ 

To take precautions, standardization organizations like NIST and governments around the world have initiated efforts to prepare for post-quantum cryptographic (PQC) landscape. The transition and adopting PQC is not only a matter of replacing the existing algorithms with the new ones, but rather it requires a careful assessment and evaluation of the system compatibility, implementation costs, deployment timelines \cite{bindal2024}. \\ 

The transition is especially harder for big organizations, and the generalized one-size-fits-all approach is not applicable since each organization's system has its own unique attributes, such as its function, user base, regulatory requirements, and operational lifetime \cite{joshi2024}. \\ 

There are numerous works that focus on inspecting the vulnerabilities of the traditional encryption methods mentioned. This paper outlines those vulnerabilities on a formal level. Additionally, to address the complex decision-making problem for organizations, we propose a machine learning-based approach designed to assist organizations in planning their quantum transitions. A model that classifies systems based on their characteristics and recommends suitable transition strategies is built. The goal of this solution is to provide a data-driven and dynamic approach to support organizations in their systems' assessments and keep the level of quantum threats to a minimum. \\

The dataset used involves 500+ systems, each has their own characteristics in terms of different features like key size, algorithm in use, expected security lifetime, and system complexity. These 500+ data points are used to train and evaluate couple of classification systems such as random forest and decision trees, to predict the most appropriate strategy from five transition categories: \\  

\begin{itemize}
    \item Immediate Replacement
    \item Hybrid Deployment
    \item Scheduled Migration 
    \item Monitor and Prepare
    \item No Action Required
\end{itemize}
\vspace{1em}

\section{Related Work - Evaluation of Quantum Security Approaches }

A major transformation in cybersecurity has been driven by the rapid growth of quantum computing, which calls for the creation and use of post-quantum cryptography (PQC) solutions to protect digital assets. PQC algorithms, implementation tactics, standardization procedures, and transition frameworks have become the subject of more and more scholarly and commercial research within the last ten years. This section surveys the existing literature, organizing the discussion into several core areas, including cryptographic primitives, standardization efforts, deployment strategies, machine learning in PQC, practical applications, and emerging interdisciplinary considerations

\subsection{Post-Quantum Cryptographic Primitives and Algorithmic Performance}

An important amount of work has been done in evaluation and optimization of PQC primitives such as key encapsulation mechanisms (KEMs) and digital signature algorithms. In constrained environments, performance of digital signature algorithms are examined by Vidaković and Miličević~\cite{vidakovic2023}, which resulted their research in finding CRYSTALS-Dilithium and SPHINCS+ to be among the most practical algorithms. Complementing this, Wang et al.~\cite{wang2024} presented efficient GPU implementations of SPHINCS+, improving performance for parallelizable devices. Hu et al.~\cite{hu2023} explored the hardware optimization of SPHINCS+ under realistic constraints, contributing to the understanding of implementation trade-offs. \\ 

In terms of key exchange, Kyber is evaluated by ~\cite{kim2023} on edge devices, showing its energy efficiency and performance advantages. Similarly, a hybrid signature scheme is suggested by Kwon et al.~\cite{kwon2024} brings together the traditional and post-quantum approaches to help a less challenging transition during the hybrid cryptographic era. CRYSTALS-Dilithium’s security in the quantum random oracle model, is examined by Jackson et al.~\cite{jackson2024}, indicating that it provides a theoretical foundations for its deployment.

\subsection{Standardization Efforts and Benchmarking}

The main element of PQC adoption is Standardization. Chen and Jordan~\cite{chen2023} offered an update on NIST's standardization work, which remains important since it established PQC protocols which are globally accepted. Lee and Kim~\cite{lee2023} proposed authentication schemes specifically customized for 5G networks, integrating quantum-safe algorithms in communication infrastructures. Bhunia et al.~\cite{bhunia2023} provided a comparative review of lightweight implementations by pointing out a benchmarking of various NIST candidates for embedded systems.  \\ 

Szymanski~\cite{szymanski2024} has developed a software-defined IoT architecture with integrated PQC capabilities to align implementation practices with secure hardware enforcement.  Hash-based signature schemes were reviewed by Fathalla and Azab~\cite{fathalla2024}, who emphasized the importance of hash functions in a post-quantum world while maintaining compatibility with existing security infrastructures.

\subsection{Transition Strategies and Frameworks for PQC Migration}

The bridging of the current systems with quantum-resilient infrastructures are important aspect of the problem, and effective migration strategies are critical to accomplish this. To come up with a system that can offer actionable guidelines for legacy systems, Hasan et al.~\cite{hasan2023} has worked on a comprehensive framework for migration of those systems, focusing mainly on security dependency analysis and organizational case studies. Aydeger et al.~\cite{aydeger2024} outlined the transition strategies from different important perspectives, such as operational and architectural. His contribution on these aspects helped detailing implementation blueprints and risk mitigation steps. \\ 

Rodriguez and Taha~\cite{rodriguez2024} evaluated quantum-safe VPNs, addressing system-level deployment challenges. Their work highlighted interoperability concerns, especially in large-scale heterogeneous networks. Park and Choi~\cite{park2023} proposed a unified PQC migration model for cloud infrastructures, advocating for modular, tenant-aware transitions. \\ 

Additionally, for one of the critical components in today's digital ecosystems, mobile apps, Xu and Ren~\cite{xu2023} dealt with a PQC-based secure update delivery. Their study revealed practical design challenges in balancing cryptographic strength with mobile device limitations. To draw the academia's attention to deficiencies in current frameworks and recommending foundational reforms, Lloyd-Jones and Manwaring~\cite{lloydjones2024} have proposed a national security perspective.

\subsection{Machine Learning for Cryptanalysis and Migration Assistance}

One of the factors that plays an increasingly important role in PQC domain is Machine learning (ML). There are numerous research on how ML can enhance cryptanalysis of PQC algorithms. In their work, Gouvêa and Pereira~\cite{gouvea2023} explored how ML can enhance it by flagging new attack surfaces previously regarded as secure. Their research's result is an important pointer that addresses the importance of adversarial learning in testing the robustness of NIST finalists. \\ 

Conversely, Cai and Ding~\cite{cai2023} focused on ML-assisted side-channel analysis, demonstrating that even post-quantum schemes are susceptible to data leakage under sophisticated probing conditions. Their work helps to highlight the need for dependency and collaboration between ML and cybersecurity communities. \\ 

Additionally, the use of ML for optimizing migration decisions and predicting system compatibility was examined by Alzahrani and Alzahrani~\cite{alzahrani2023}. They proposed a cybersecurity maturity model that leverages ML to assess organizational readiness for PQC adoption. Gouvea and Pereira’s study presented a research roadmap for integrating AI into PQC-decision making frameworks, later complemented the Alzahrani's approach. 

\subsection{System-Level and Protocol-Specific Applications}
Numerous studies have explored the integration of PQC into full-stack systems. Garcia et al.~\cite{garcia2024} analyze the incorporation of post-quantum algorithms into Transport Layer Security (TLS), highlighting the associated latency and throughput trade-offs. For end-to-end secure messaging, which is a significantly growing method in secure communication services,   Bhargavan et al.~\cite{bhargavan2024} offered a formal verification of the PQXDH protocol, establishing rigorous proofs for introducing post-quantum to those services. \\ 

Verchyk and Seplveda~\cite{verchyk2023}, who proposed functional deployment strategies for secure communication infrastructures, using quantum-resistant algorithms investigated identity-based encryption augmented. Nguyen and Miyazaki~\cite{nguyen2023} proposed hybrid key exchange mechanisms that merge lattice-based cryptography with elliptic curve methods to ensure transitional security in different environments. \\ 

The need for privacy and authentication in IoT sytems is addressed by the work of Mansoor et al.~\cite{mansoor2024}, which applied PQC to IoT ecosystems. For IoT nodes that have resource constraints, their work emphasized that lightweight implementations such as Kyber512 and Falcon-512 are more appropriate. Their research results align with those of Bhunia et al.~\cite{bhunia2023}, which  strengthens and validates the Kyber's advantages as a candidate for low-power systems.

\subsection{Legal, Ethical, and Policy Considerations}

Although technical feasibility has been the primary focus of existing literature, recent researchs have begun to examine the wider implications of post-quantum cryptography (PQC). Compliance with regional legal structures continues to complicate the regulatory landscape surrounding quantum innovations. Dang~\cite{dang2024} examined these regulatory issues and underscored the need for adaptable policy mechanisms capable of responding to quantum-era threats. \\

Alao et al.~\cite{alao2024} looked into how putting off the adoption of post-quantum cryptography could shake financial stability and weaken governance. They cautioned that delays might damage investor trust and potentially cause broader economic issues. Similarly, Lloyd-Jones and Manwaring\cite{lloydjones2024} highlight the role of weak regulatory frameworks in increasing these risks, arguing that many current cybersecurity strategies fail to address the unique challenges posed by quantum technologies. \\

Arigbabu et al.~\cite{arigbabu2024} broadened the scope of discussion to the healthcare sector, examining how AI-driven data governance could be affected by quantum security risks. They emphasized the urgency for healthcare systems handling sensitive patient data to adopt post-quantum cryptography in order to mitigate potential future risks.

\subsection{PQC in Blockchain, Edge, and Emerging Technologies}

The intersection of PQC and blockchain is gaining speed. Marchsreiter~\cite{marchsreiter2025} investigated PQC-based blockchain signatures on embedded systems, highlighting effective key recovery strategies. This research highlights the potential of quantum-secure decentralized applications, especially in critical sectors such as financial services and logistics. \\

Garg and Garg~\cite{garg2025} offered a thorough overview of post- quantum cryptography (PQC) and quantum key distribution (QKD), evaluating how these technologies could work together to protect future communication networks. They suggested that while quantum key distribution (QKD) offers high security, post-quantum cryptography (PQC) tends to be more practical and scalable for broader implementation. \\

In the field of edge computing, Kim et al.\cite{kim2023} and Szymanski\cite{szymanski2024} suggested the benefits of using deterministic, hardware-based systems. Their work showed that post-quantum cryptography (PQC) can be integrated into low-latency environments without sacrificing performance, reinforcing the idea that PQC solutions should be customized to be tailored to the specific needs of each environment.

\subsection{A Holistic Perspective on Quantum-Resilient Infrastructures}

Recent studies show a growing trend of treating post-quantum cryptography (PQC) as a foundational element in system architecture. For example, Baseri et al.~\cite{baseri2024} offered a broad perspective on quantum-secure networking, recommending for layered defense strategies that combine cryptographic techniques with practical security measures. \\

Similarly, Dang~\cite{dang2024} and Garg and Garg~\cite{garg2025} pointed out the importance of cross-disciplinary collaboration, which brings together legal, technical, and engineering fields, to tackle the complex issues involved in implementing PQC. Building on this, Arigbabu et al.\cite{arigbabu2024} and Alao et al.\cite{alao2024} examined the wider societal and ethical impacts of bringing PQC into modern digital systems. \\

Meanwhile, Nguyen and Miyazaki~\cite{nguyen2023}, along with Kwon et al.~\cite{kwon2024}, introduced hybrid transition models that support the increasingly accepted view that the move to PQC will be gradual and layered. Their findings suggest that PQC adoption is not merely a technical shift—it represents a fundamental transformation of the entire digital ecosystem.

\subsection{Summary and Research Gap Analysis}

Earlier studies have laid a strong foundation for PQC research, covering main areas like algorithm efficiency, real-world implementation, formal verification, migration strategies, machine learning integration, and broader socio-technical concerns. Still, there are several important gaps that have yet to be filled. \\

First, although individual algorithm performance on specific hardware has been widely evaluated, comprehensive performance analyses spanning cloud, edge, and mobile platforms are limited. Secondly, little attention has been paid to the incorporation of machine learning into migration strategies, especially in the realms of adaptive threat modeling and anticipatory deployment planning. \\

Third, interdisciplinary cooperation involving policymakers, industry stakeholders, and cryptographic researchers remains at an early stage, despite its critical role in achieving sustainable long-term outcomes. Finally , even though hybrid approaches to PQC adoption are becoming increasingly common, empirical investigations into their real-world practicality and effectiveness, especially under adversarial conditions, are scarce. \\

To ensure a smooth, secure, and ethical transition to quantum-resilient systems, future research needs to focus on closing these gaps.

\section{Proposed Solution - Data-Driven Framework for Cryptographic Migration}
The proposed solution, a decision framework designed to recommend cryptographic migration strategies for organizations transitioning to systems that are quantum-resilient. The framework, which is referred as the Quantum Transition Strategy Recommendation Framework (QTSRF) maps the characteristics of the existing systems to one of the several actionable transition strategies. 

\subsection{Dataset Construction}
As the studies are newly emerging in the field, finding a present dataset for Post-Quantum Cryptography is a hassle. A synthetic dataset created in the light of a variety of white papers from industry and academic papers to employ the most generalistic approach to propose a solution. The features that are utilized in the dataset are: 
\begin{itemize}
    \item System Type
    \item Security Lifetime 
    \item Cryptographic Method/Algorithm
    \item Key Size
    \item System Complexity
    \item Integration Complexity 
    \item Data Sensitivity 
    \item Recommended Strategy
\end{itemize}

\vspace{1em}

The dataset comprises of 500+ records, each describing a digital system through carefully selected attributes: \\

\subsubsection{Security Lifetime Requirements}
This field indicates how many years a system's cryptography must remain secure. Mosca (2018) proposes an light mathematical equation for understanding the urgency of cryptographic migration. We blended this concept into our solution in threshold setting and migration strategy categories. \\ 

In threshold setting, if the security lifetime exceeds a threshold (this is typically set to 10) in systems that use algorithms like RCA or ECC - whose security is known to be vulnerable under quantum threat estimates - those systems are flagged as high-risk. This flagging decision is informed by Mosca's estimate that RSA-2048 might start becoming vulnerable in the 15-20 years window. \\ 

\subsubsection{Cryptographic Method and Key Size: Hybrid Strategy Justification}
The two parameters, crypto\_method and key\_size are essential to identify whether the encryption mechanisms are quantum-vulnerable, quantum-neutral, or quantum-resistant. These two parameters join the proposed decision mechanism to express a system's current level of cryptographic strength and resilience to quantum threats. \\

In our dataset, systems that: \\
\begin{itemize}
    \item Use RSA or ECC as primary cryptographic method
    \item Have key sizes $\leq$ 2048 (for RSA) or $\leq$ 256 (for ECC)
    \item And also have high integration\_complexity (with scores of 4 or 5) or moderate-to-high system complexity 
\end{itemize}

\vspace{1em}

are assigned the immediate\_hybrid strategy. Although this shows that full cryptographic overhaul may not be feasible in the short term, reflects the urgency of strengthening their cryptographic posture. This approach mirrors the existing studies by mitigating the quantum threat incrementally without destabilizing existing operations. \\

\subsubsection{System \& Integration Complexity Correlations}
As ENISA (2021) stated, the systems that protect highly sensitive data over long periods of time are more likely to have a layered security requirements, which naturally make their architecture more complex. The system complexity and integration complexity fields in our dataset quantifies the technical and practical difficulties of updating these complex environments. \\

To quantify those, our dataset has a scaling mechanism rating from 1 to 5 for both system\_complexity and integration\_complexity. A simple architecture and minimal integration challenges are rated as 1, and high complex system with extensive legacy dependencies and challenging integration scenarios are identified as rate 5. \\  

The developed framework assigns more strict migration plans to systems with higher complexity rates. For instance, a system that has a security\_lifetime of greater than 10 years that also scores 4 or 5 in both system\_complexity and integration\_complexity is typically flagged for an immediate\_hybrid or scheduled\_transition strategy. \\

\subsubsection{Data Sensitivity - Prioritizing Critical Systems}
In our dataset, data\_sensitivity attribute serves as a strategic risk indicator, which quantifies the level of criticality, confidentiality, and impact of the data the system handles. Our dataset also has a rating mechanism in the range 1-5 for data\_sensitivity, where: \\

\begin{itemize}
    \item 1: Low Sensitivity
    \item 2-3: Medium Sensitivity
    \item 4:5 High Sensitivity 
\end{itemize}

\vspace{1em}

We embed this logic into our framework by flagging systems that has data\_sensitivity $\geq$ 4 as high-priority. immediate\_hybrid or scheduled\_transition depending on other factors like integration complexity and crypto method. Generally, payment\_processing, secure\_messaging, and certificate\_authorities often fall into this category. 

\subsection{Analytical Framework}
We formalize the relationship between the cryptographic methods and recommended strategies by introducing a risk-based analytical formula. $R(s,t)$ represents the quantum-risk of a system  $s$ over time  $t$, which can be represented as: 

\vspace{1em}

\begin{equation}
R(s,t) = V(m,k) \cdot S(d) \cdot P(t)
\end{equation}
\\

Where: 
\begin{itemize}
    \item $V(m,k)$ is the vulnerability function based on crypographic method $m$ and key size $k$. 
    \item $S(d)$ is the sensitivity scaling factor based on data sensitivity $d$. 
    \item $P(t)$ is the probability function of quantum computing capability reaching the necessary threshold by time $t$.  
\end{itemize}

\subsection{Implementation Overview}
As mentioned above, the dataset is structured such that it consists of system\_type, security\_lifetime, crypto\_method, key\_size, system\_complexity, integration\_complexity, data\_sensitivity features and recommended\_strategy as target feature. \\ 

\subsubsection{Preprocessing}
In terms of preprocessing, to be able to interpret non-numerical features, the categorical fields like system\_type and crypto\_method are one-hot encoded using OneHotEncoder, which resulted in binary feature vectors that allowed the model to interpret those features. \\ 

To ensure the uniformity of the input for the model, a stacked array of normalized numerical features and encoded categorical features created as a final feature matrix to be fed into the model. \\ 

\subsubsection{Train-Test Split} To make sure that model performance metrics are not biased by unbalanced labels, the processed data is split into 70\% training and 30\% test sets. \\

\subsubsection{Model Training} For comparison purposes, two models are trained: \\
\begin{itemize}
    \item Decision Tree Classifier
    \begin{itemize}
        \item To prioritize the interpretability, a shallow tree with a maximum depth of 5 is trained. 
    \end{itemize}
    \item Random Forest Classifier
    \begin{itemize}
        \item For the sake of generalizability and accuracy, an ensemble that consists of 100 decision trees is trained. Random Forest is chosen for its resilience to noise and capability to understand complex inter-dependencies and relations between features.    
    \end{itemize}
\end{itemize} 

\vspace{1em}

\subsubsection{Model Evaluation}
Decision Tree and Random Forest models are evaluated using:  \\
\begin{itemize}
    \item Classification reports to detail the precision, recall, F1-score.  
    \item Confusion matrices 
    \item Feature importance analysis to rank the features and highlight the top-contributing ones. 
\end{itemize} 

\vspace{1em}

\subsubsection{Outlined Decision Making Process}
The decision rules of the Decision Tree are outlined in a readable format which is easy to track to introduce transparency. This enables further technical inspection by technical teams like security teams within organizations if need be. \\ 

\subsubsection{Prediction Function \& Output}
The prediction function accepts an dictionary input matching the structure of the dataset, encodes the input and predicts the most likely strategy using the Random Forest model. The output of the model contains the \textbf{recommended strategy}, \textbf{model confidence}, and \textbf{top 3 alternatives with their probabilities}. 

\subsection{Design Objectives}
\begin{itemize}
    \item Interpretable
    \begin{itemize}
        \item Decision Tree model enables the proposed framework to be transparent by outlining the decision rules and process. 
        \item Designed in a way that both technical and non-technical stakeholders can understand and track the decision process easily. 
    \end{itemize}
    \item Robust
    \begin{itemize}
        \item Random Forest model introduces a robustness to framework by introducing robusness against systems that are unseen and newly introduced to the model,  
    \end{itemize}
    \item Scalable
    \begin{itemize}
        \item The framework is implemented in a modular way, enabling the addition new features to the system which can enhance the prediction of the system by introducing different dimensions. 
    \end{itemize}
    \item Practical 
    \begin{itemize}
        \item The proposed framework enables organizations to assess their quantum-readiness. 
    \end{itemize}
\end{itemize}

\section{Analysis of the Solution}
\subsection{Dataset Analysis}
There are various approaches for creating logical synthetic data. By establishing the fundamental patterns of our data based on the features of the systems and their effects on the quantum-readiness of the system according to the recent work on security principles and expert knowledge mentioned in the previous sections, we developed a specialized synthetic data generation methodology that create logically consistent cryptographic system examples based on those fundamentals. \\ 

To capture the theoretical and practical relationships between cryptographic systems and quantum vulnerability, we defined specific rule sets by following the principles from the related work. These rule sets led us to the formula we built that is mentioned in the Analytical Framework section previously. After implementing the base of the formula, the system-specific incorporated by defining domain-specific constraints for different system types. For instance, higher data-sensitivity requirements for healthcare and military systems, lower complexity thresholds for IoT devices, and appropriate integration complexity factors for embedded systems are implemented. \\ 

As a result of this dataset synthesis, we acquired a balanced dataset containing 1205 records with 241 samples per strategy class, to prevent the bias in favor of any strategy class. A validation process is conducted to evaluate if the synthetic data points follow the established fundamental logical cryptographic relationships. This validation process confirmed that 99.4\% of those data points maintained the relationships, with only minor inconsistencies. \\

\begin{figure}[H]
\centering
\includegraphics[width=3.5in]{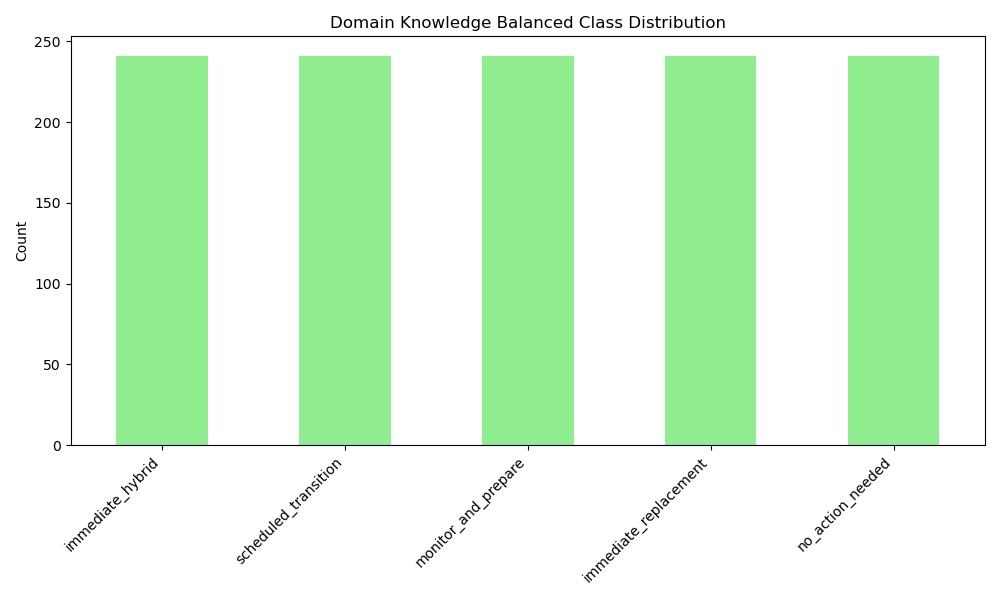}
\caption{Distribution of each strategy class in the dataset}
\label{fig_1}
\end{figure}

\subsection{Feature Importance Analysis}
When the output of the Random Forest model is inspected, it is seen that critical insights about the decision factors are revealed: \\  
\begin{enumerate}
    \item \textbf{Temporal Security Requirements}: Security lifetime is identified as the most influential feature (24.3\%), indicating that the operational duration has a significant impact in transition strategy. 
    \item \textbf{Cryptographic Strength}: Key size is identified as the second most important feature (20.9\%), indicating that strength of the existing cryptographic implementations affects transition strategy. 
    \item \textbf{Cryptographic Algorithm}: It is observed that specific cryptographic methods (with RSA being the most significant with 8.9\%) surpassed even the factors like system complexity and integration complexity. 
    \item \textbf{Implementation Factors}: System complexity and integration complexity are identified as secondary factors by 7.4\% and 6.0\% respectively. 
\end{enumerate}

\begin{figure}[H]
\centering
\includegraphics[width=3.5in]{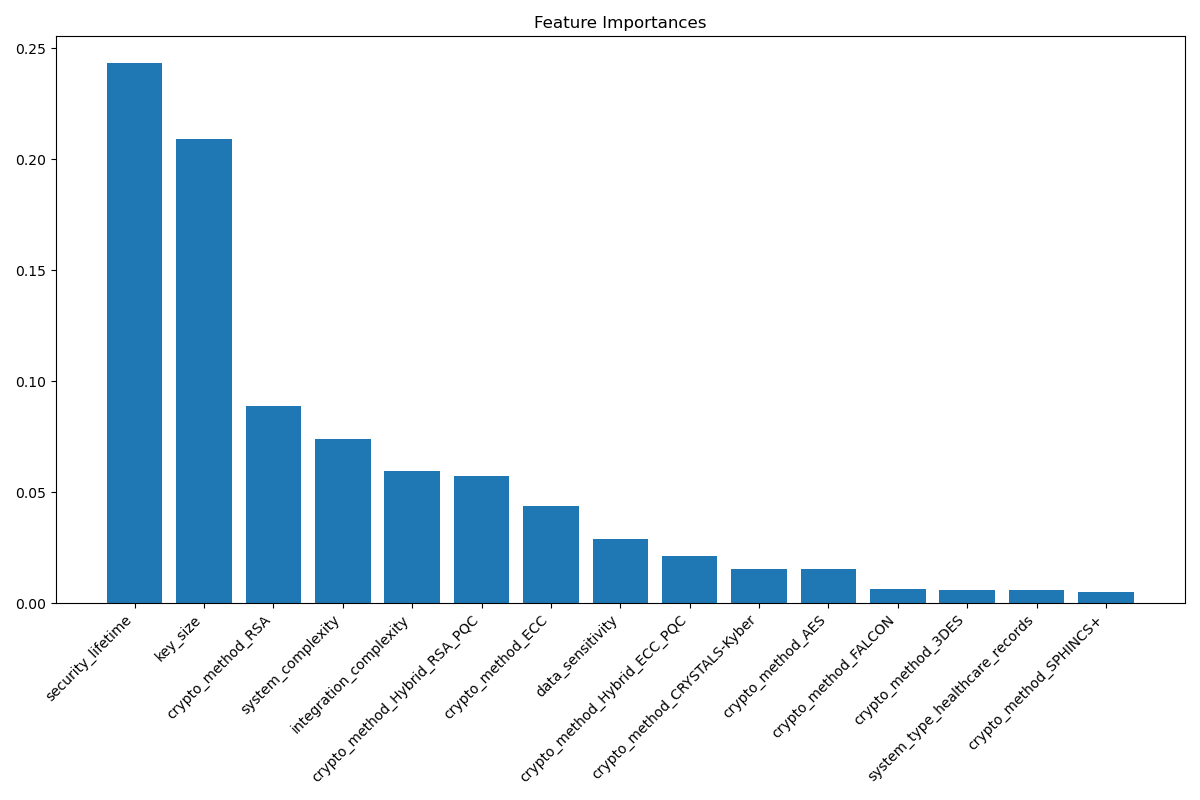}
\caption{Feature Importances}
\label{fig_1}
\end{figure}

When the heatmap analysis of the relationship between the cryptographic methods and strategies support the analytical model that is introduced as analytical framework in the proposed solution section. 
\vspace{1em}

For instance, our vulnerability function, $V(m,k)$, clearly differentiates between post-quantum methods (CRYSTALS-Kyber, CRYSTALS-Dilithium, FALCON) with near-zero vulnerability values and legacy methods (3DES, RSA-1024) with high vulnerability values. This is proven by the 95-100\% strong correlation results between post-quantum methods and the "no\_action\_needed" strategy. 
\vspace{1em}

Cryptographic methods and strategy relationship heatmap provides a significant results on the differentiation between them: 

\begin{enumerate}
    \item \textbf{Post-Quantum Confidence:} The correlation between post-quantum algorithms and the "np\_action\_needed" plan (95-100\%) is near to the perfection. 
    \item \textbf{Hybrid Approach vs Transitional Solution:} monitor\_and\_prepare plan is found as best suitable plan for hybrid approaches like Hybrid\_RSA\_PQC and Hybrid\_ECC\_PQC, indicating rather than providing a long-term security measures, those methods are playing their role by providing transitional solutions. 
    \item \textbf{RSA Algorithm Vulnerability Spectrum:} It is shown that RSA has a balanced distribution across different transition plan types - immmediate\_replacement 35\%, immediate\_hybrid 32\%, scheduled\_transition 29\% - indicating that its position in the vulnerability spectrum is based on other factors like key\_size, and other complexities. 
    \item \textbf{ECC Transition Requirements:} The distribution of ECC across various transition strategies indicates that it provides better security than RSA, however not resilient enough against quantum attacks, therefore will eventually require quantum-safe replacements. 
\end{enumerate}

When the cryptographic methods vs strategies heatmap is examined on a broader perspective, the results can be interpreted as higher-complexity systems tend toward hybrid approaches rather than immediate replacement due to the high risks caused by migrating the complex systems and replacing cryptography in those complex environments. 

\begin{figure}[H]
\centering
\includegraphics[width=4in]{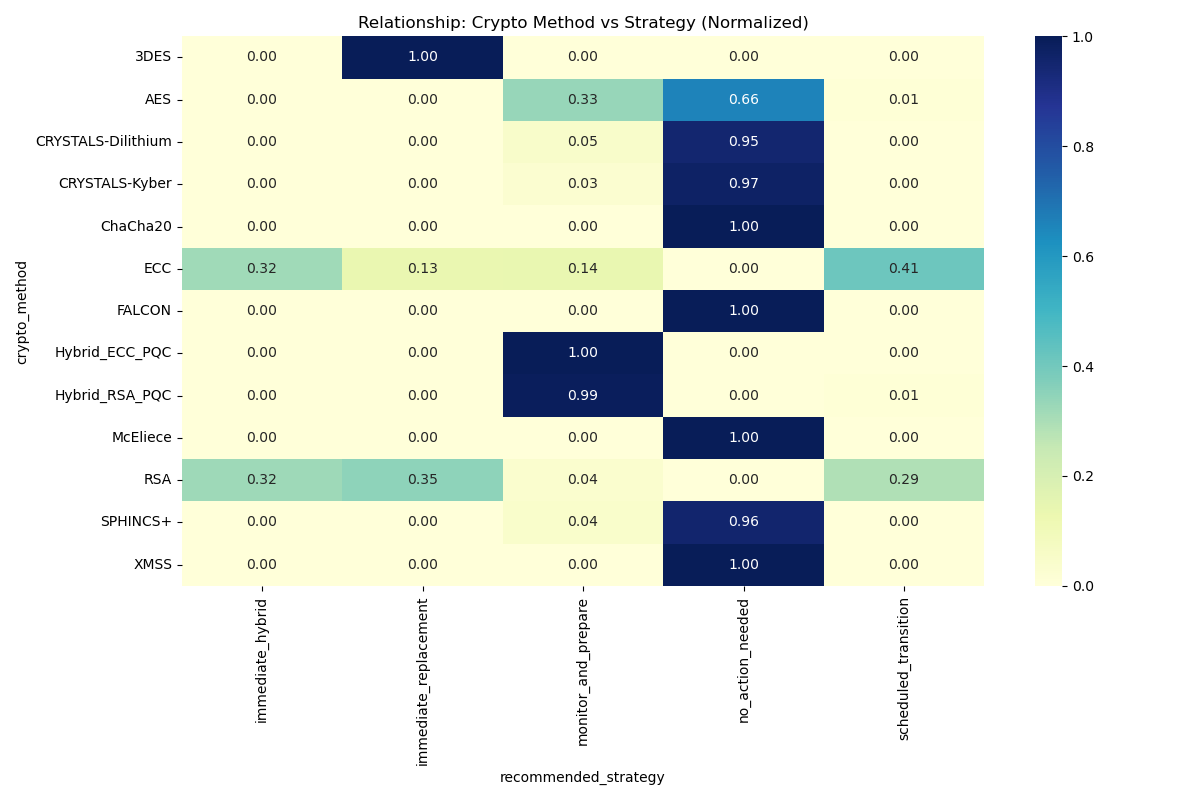}
\caption{Cryptographic Method - Recommended Strategy}
\label{fig_1}
\end{figure}

Additionally, when the system types and their correlation to vulnerability is analyzed, interesting results are observed. To quantify this correlation, a vulnerability scoring methodology that converts categorical strategy recommendations into numerical vulnerability index is developed. A scale from 1 to 5 is created to map each strategy to a point in the scale, with higher values indicating greater vulnerability and more urgent transition requirements. This approach enabled developed framework to quantify each system type's relative susceptibility to quantum threats. Standard deviation values are also computed to assess the variability of vulnerability within each system type category. \\ 

For instance, system types like payment sytems (3.70), military communications (3.41), and healthcare records (3.36) have highest average vulnerability scores, indicating they require the most urgent quantum-safe transitions. It is also observed that while many system types have relatively balanced strategy distributions, some of the systems like weather forecasting and wireless networks tend to fall under the "no action needed" and "scheduled transition" strategies. \\

\begin{figure}[H]
\centering
\includegraphics[width=4in]{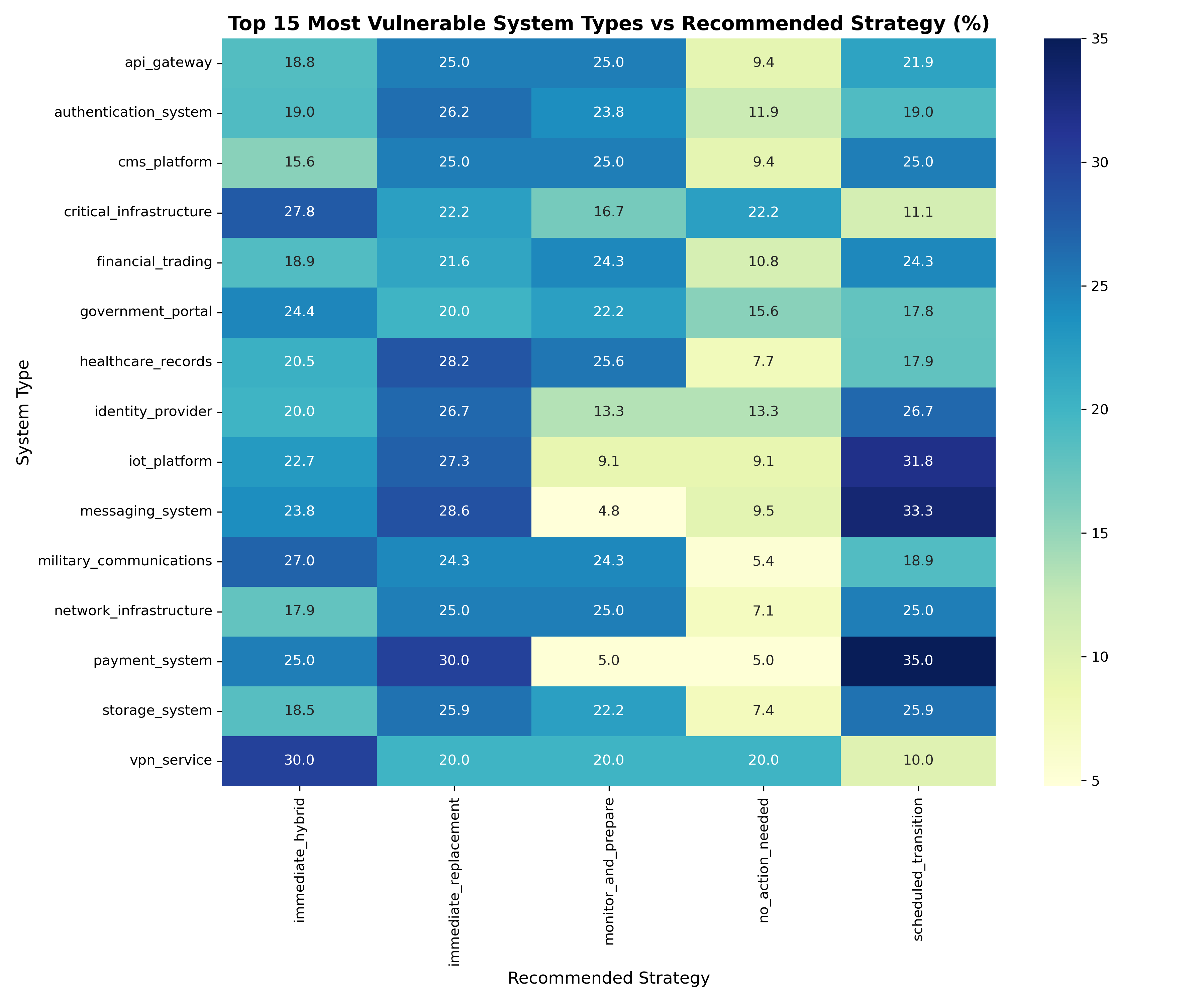}
\caption{System Type - Recommended Strategy}
\label{fig_1}
\end{figure}

\subsection{Model Performance}
\subsubsection{Decision Tree}
The results of the Decision Tree model are respectable, but noticeably lower than Random Forest. Its overall accuracy is identified as 81\%. F1 scores, like 0.96 for immediate\_replacement, but 0.62 for monitor\_and\_prepare shows that it performs well on some classes, and struggles with others. Its cross-validation score of 79.5\% $\pm$ 10.14\% shows more inconsistency. \\ 

\subsubsection{Random Forest}
Overall accuracy of the Random Forest Model is 96\%. It also has a very balanced per-class performance indicated by the F1-scores ranging from 0.92 to 1.00. Additionally, its cross-validation of 91.78\% $\pm$ 3.48\% proves that it has good consistency and stability across different data splits. \\

\par The much higher standard deviation for the Decision Tree indicates that it is not adaptive to the new systems' data introduced to it. Therefore, Random Forest's ensemble approach is more robust and handles that better. Thus, in terms of real-world reliability, Random Forest is more applicable considering the model will encounter new systems in a real-world scenario. \\ 

\par However, it is important to highlight that there is a tradeoff between the model's complexity and its performance. Although the performance gap between two models are significant and obvious (96\% vs 81\%), in a real-world deployment scenario, deciding which model to employ would depend on the organization's purposes considering whether interpretability or accuracy is more critical for their quantum-safe transitioning planning. Because, Decision Tree model is significantly simpler and more interpretable (single tree vs 100 trees).  

\begin{figure}[H]
\centering
\includegraphics[width=3.5in]{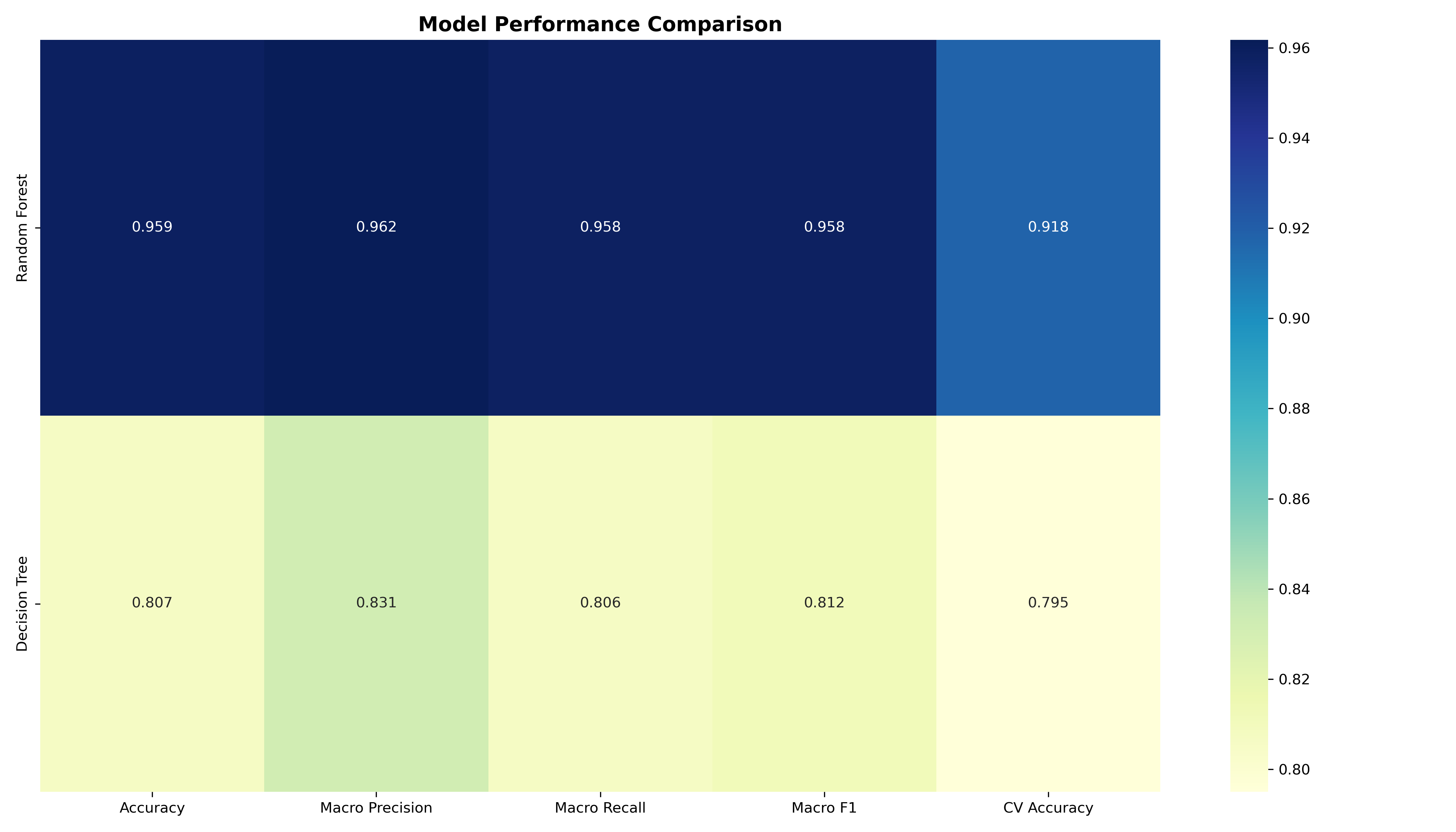}
\caption{Model Performance Heatmap}
\label{fig_1}
\end{figure}

In the confusion matrices, the Random Forest model exhibits a diagonal dominance, which clearly indicates high accuracy across all classes. The "immediate\_replacement" and "no\_action\_needed" classes are perfectly classified with zero misclassifications. \\

For the sake of comparison, confusion matrix gives meaningful insights: 
\begin{itemize}
    \item The decision boundaries between similar strategies are handled better by the Random Forest's ensemble approach. 
    \item There are some classes that both models struggle most to distinguish between. "monitor\_and\_prepare" and "scheduled\_transition" are two of them, suggesting they share similar characteristics. 
    \item There are classes that are super straightforward for the models to distinguish between, such as "immediate\_replacement" and "no\_action\_needed", indicating that they have the most distinctive features. 
\end{itemize}

\begin{figure}[H]
\centering
\includegraphics[width=3.5in]{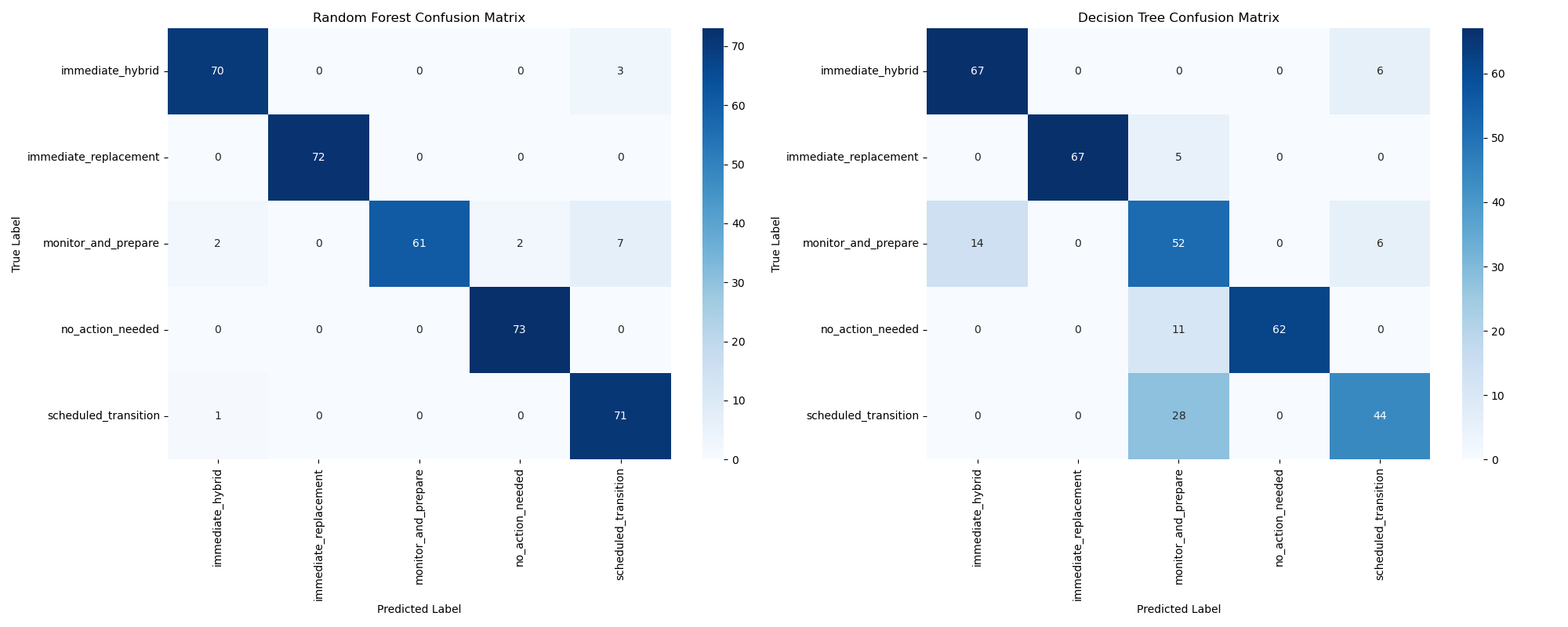}
\caption{Confusion Matrices}
\label{fig_1}
\end{figure}

\begin{figure}[H]
\centering
\includegraphics[width=3.5in]{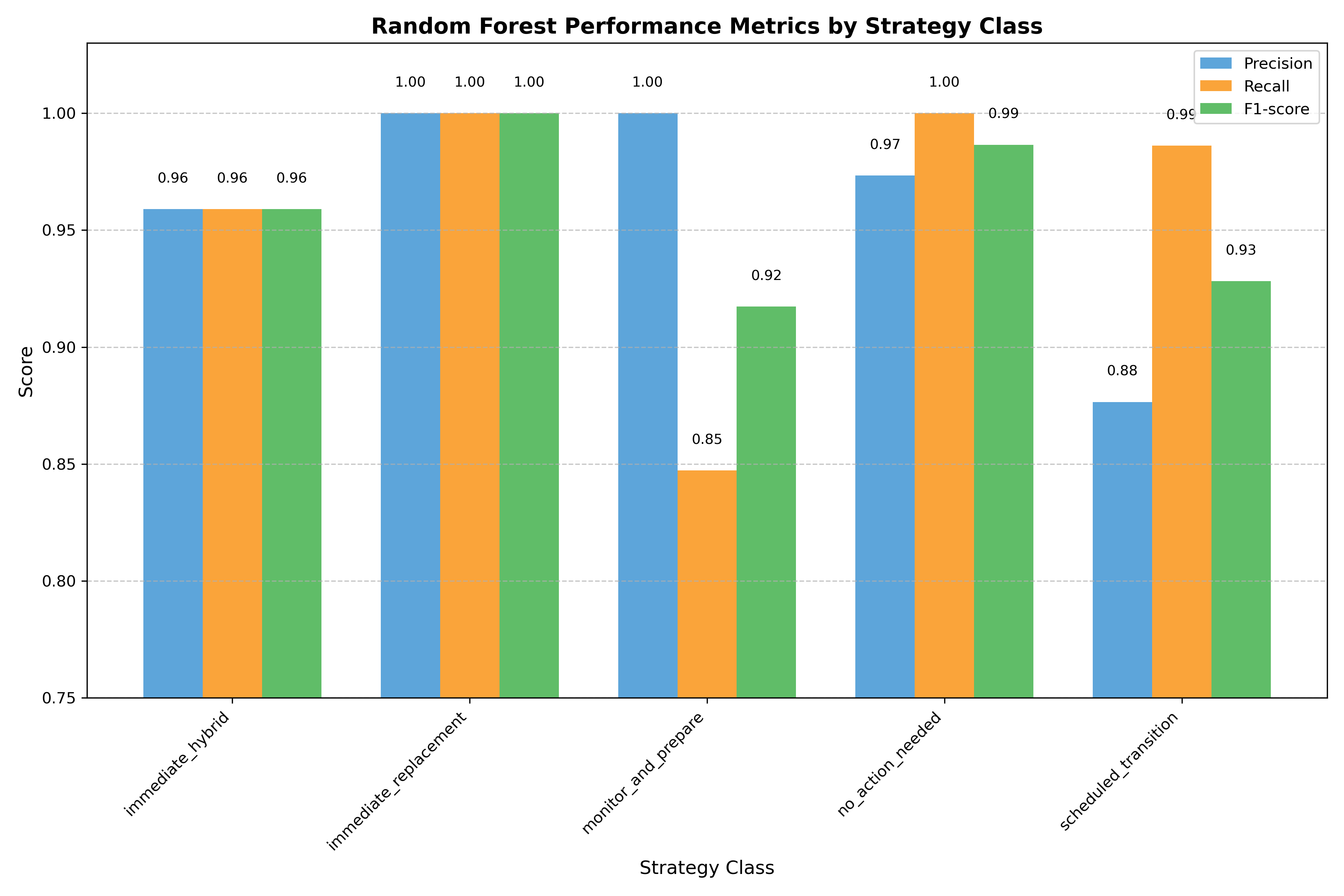}
\caption{Random Forest Model - Class Performance}
\label{fig_1}
\end{figure}

\section{Conclusion and Future Work}
Developments in quantum field is causing threats against traditional networks and cryptographic methods. Hence, both public and private organizations like universities, research institutes, private companies are conducting research on this topic to develop either defense or transition mechanisms. However, it is challenging to come up with a single solution that covers all of the problems across different specific systems as they have unique setups. Therefore, there is an urgent need for a systematic approach to guide the organizations by providing solutions specific to them, or at least guide them to take calculated precautions against threats until a concrete solution is developed in the industry or academia. Additionally, this systematic approach needs to balance the security requirements with implementation realities of unique systems. \\   

Due to the lack of data related to post-quantum cryptography and quantum attacks, the solution proposed in this paper employs a domain knowledge-based synthetic data generation approach by taking the base knowledge and established cryptographic security principles on the features that plays an important role in systems' vulnerability against quantum threats, create logically consistent examples that accurately reflect those principles, ensuring that the model learns meaningful patterns rather than statistical artifacts. Random Forest model is selected to predict appropriate quantum-safe transition strategies, due to its ability to handle complex non-linear relationships between features, robustness against overfitting, and superior performance in capturing the multifaceted decision boundaries between different transition strategies as demonstrated by its 96\% accuracy compared to simpler alternatives. Due to the nature of the attributes in the created dataset, the decision making process integrates both algorithm characteristics like encryption method, key size, and also implementation contexts such as system complexity and data sensitivity. \\ 

The proposed solution contributes to research in this field by providing a logically built balanced and domain-specific dataset which relies on the established principles about quantum threats by the previous research. Thus it lights the way for future work by providing a dataset that has the characteristics of the domain. The solution achieves 96\% accuracy in recommending appropriate transition strategies. Additionally, it identifies the key factors like security lifetime, key size etc. and analyzes their contribution to the decision mechanism. The proposed work specifically highlights and demonstrates the relationship between cryptographic methods and appropriate transition strategies. \\ 

Due to the newly emerging nature of the field, the developed solution is open to progressive refinements in the future as quantum computing research advances and post-quantum cryptography implementation matures, which can transform the model from only being domain expertise to a hybrid system that leverages both theoretical knowledge and empirical evidence from the field. With the light of more data and standardized post-quantum algorithms, the model can be refined to make customized recommendations and organization-specific risk tolarence profiles can be developed. Since this is a fast-growing field, a mechanism to validate the decision making process of the model can be developed to make it more robust to the latest advancement in the quantum field. \\ 

As the proposed solution applies a domain-based theoretical knowledge in building a framework that organizations can use for their security assessments and their use-cases, it has a potential practical value. Security practitioners in organizations can integrate the framework into their systems and take advantage of the model's insights in the enterprise security planning processes. Therefore, the framework has potential impact on organization-wide cryptographic governance and risk management.

\end{document}